\documentclass{article}

\newcommand{\text}{\rm }

\begin{document}

\begin{center}
{\large \bf Influence of symmetry breaking on the fluctuation properties of spectra}\\

\vspace{0.5 in}
{\bf  A  Abd El-Hady}, {\bf A Y Abul-Magd}  
{\bf  and M H Simbel}\\
Faculty of Science, Zagazig University, Zagazig, Egypt \\
\vspace{0.5 in}

\end{center}

\begin{abstract}

We study the effect of gradual symmetry breaking in a non-integrable system
on the level fluctuation statistics. \ We consider the case when the
symmetry is represented by a quantum number that takes one of two possible
values, so that the unperturbed system has a spectrum composed of two
independent sequences. When symmetry-breaking perturbation is represented by
a random matrix with an adjustable strength, the shape of the spectrum
monotonously evolves towards the Wigner distribution as the strength
parameter increases. This contradicts the observed behaviour of the acoustic
resonance spectra in quartz blocks during the breaking of a point-group
symmetry that has two eigenvalues, where the system changes in the beginning towards the Poisson statistics
then turns back to the GOE statistics. This behaviour is explained by assuming
that the symmetry breaking perturbation removes the degeneracy of a limited number of levels, thus creating a third chaotic sequence. As symmetry breaking increases, the new sequence grows at the expense of the initial pair until it overwhelms the whole
spectrum when the symmetry completely disappears. The calculated spacing
distribution and spectral rigidity are able to describe the evolution of the
observed acoustic resonance spectra.

\newpage
\end{abstract}

\section{Introduction}

The statistical theory of spectra \cite{phys.rep} provides an appropriate
method for examining the symmetry properties of quantum systems. \ Level
statistics such as the nearest-neighbour spacing (NNS) distribution and the
spectral rigidity $\Delta _{3}$ are defined for a pure sequence of levels
that have same quantum numbers. Usually, it is taken as granted that in a
classically integrable system, which has as many integrals of motion (or
quantum numbers) as the number of degrees of freedom, the levels are
uncorrelated and so have a Poissonian NNS distribution. \ The pure level
sequence in a time-reversal-invariant quantum system whose classical
counterpart is chaotic, is successfully represented by a Gaussian orthogonal
ensemble (GOE) of random matrices. \ A deviation of level statistics from
the predictions of the GOE may be taken as an evidence that the sequence is
not pure because of the existence of a hidden symmetry. \ The statistical
theory of spectra has also been extensively used in the investigation of
symmetry breaking because the destruction of a quantum number has a dramatic
effect on the spectral fluctuations. \ Impressive are the studies of isospin
mixing in light nuclei \cite{harney}. \ Shriner \textit{et al}.\cite{shriner}
measured a ''complete'' spectrum of low-lying states of $^{26}$Al involving
states with isospins $T=0$ and 1 throughout the energy range covered by the
data. \ These data offered a testing ground for studying the influence of
isospin-symmetry breaking on the fluctuation properties of energy spectra
(see, e.g., \cite{paar, guhr, hussein, simbel}). \ However, the limited data
available in $^{26}$Al (142 levels) precluded a definite conclusion. \ In a
recent experiment with monocrystalline quartz blocks, Ellegaard \textit{et
al.} \cite{ellegaard} measured about 1400 well-resolved acoustic resonances.
\ The properties of quartz allowed these authors to measure the gradual
break-up of a point-group symmetry, which is statistically fully equivalent
to the breaking of a quantum number like isospin. \ The spectra are obtained
by externally tuning the symmetry breaking \cite{ellegaard} , allowing a
statistically significant investigation of the whole transition.

In the present paper we consider the effect of gradual symmetry breaking on
the fluctuation properties of energy levels. When the symmetry is full the
spectrum is divided into independent sequences, each corresponding to one of
the eigenvalues of the symmetry operator. \ We consider two possible
scenarios for breaking the symmetry. \ The first is provided by the
Guhr-Weidenm\"{u}ller model \cite{guhr}, which assumes that the perturbation acts equally on all states and that the matrix elements between states of different symmetry eigenvalues are random numbers
of equal variances. \ In the second scenario, the perturbation acts differently on the unperturbed states and the symmetry breaking interaction \ mixes a limited number of the degenerate eigenstates that
belong to different symmetry representations, and this number grows as we
increase symmetry breaking. \ We are thus creating a new sequence of levels
corresponding to states with no symmetry at the expense of the initial
sequences. \ Consequently, as the fractional density of the no-symmetry
sequence increases, the $\Delta _{3}$ statistics of the total spectrum first
increases, moving in the direction of the prediction of the Poisson
statistics until the densities of all the sequences become equal, and then
decreases to reach asymptotically the value given by the GOE. \ Section 3
shows that the behaviour described by the second scenario is indeed present
in the data of Ellegaard \textit{et al}. \cite{ellegaard}. \ The summary and
conclusions of this work are given in Sect. 4.

\section{Symmetry breaking in a chaotic system}

We consider a chaotic system that conserves a given symmetry. \ In such a
system, the Hamiltonian assumes a block-diagonal form \cite{phys.rep}: 
\begin{equation}
H_{S}={\text{diag}}\left( H_{1},H_{2},\cdots ,H_{n}\right) \text{,}
\label{diag}
\end{equation}
\ where each sub-block corresponds to states of certain quantum numbers. \ Thus, when
the symmetry is conserved, the levels are divided into a number of ''pure''
sequences, each described by a pure GOE. \ However, the statistical
properties of the total spectrum that combines these sequences are no longer
of the GOE type. \ In the limit of infinite number of sequences combined,
the spectrum obeys the Poisson statistics. \ For a finite number of sequences,
the fluctuation properties are intermediate between the Poisson and GOE
statistics. \ In the following, we shall restrict our consideration to the
case when the symmetry under investigation is represented by an operator
that has two possible eigenvalues, such as parity, or isospin 0 and 1 as
considered in Refs \cite{shriner, paar, guhr, hussein, simbel}.

The NNS distribution of a spectrum resulting from a random superposition of $%
n$ independent sequences is calculated, e.g., by Berry and Robnik \cite
{berry} and in Mehta's book \cite{mehta}. \ If the level density of the $i$%
th sequence is $\rho _{i}$, and if the NNS distribution of levels of this
sequence is $P_{i}(x_{i})$, where $x_{i}=f_{i}s$, $f_{i}=\rho _{i}/\Sigma
\rho _{i},$ and $s$ is the NNS normalised to a unit mean, then the NNS
distribution of the mixed sequence is given by 
\begin{equation}
P(s)\!=\!E(s)\left\{ \sum_{i}f_{i}^{2}\frac{P_{i}\left( f_{i}s\right) }{%
E_{i}\left( f_{i}s\right) }\!+\!\left[ \sum_{i}f_{i}\frac{1-W_{i}(f_{i}s)}{%
E_{i}(f_{i}s)}\right] ^{2}\!\!-\!\!\sum_{i}\left[ f_{i}\frac{1-W_{i}(f_{i}s)}{%
E_{i}(f_{i}s)}\right] ^{2}\right\} ,
\label{mixedseq}
\end{equation}
where $E(s)=\prod_{i=1}^{n}E_{i}(f_{i}s), $ $E_{i}(x_{i})=\int_{x_{i}}^{%
\infty }\left[ 1-W_{i}\left( x\right) \right] dx$, and $W_{i}\left(
x_{i}\right) =\int_{0}^{x_{i}}P_{i}(x)dx.$ \ In particular, if all the $n$
individual sequences have the same level densities, so that $f_{i}=1/n$, and
if the NNS distribution in each is a Wigner distribution 
\begin{equation}
P_{i}(x_{i})=\frac{\pi }{2}x_{i}e^{-\frac{\pi }{4}x_{i}^{2}}
\label{wigner}
\end{equation}
which is a good approximation for the NNS distribution of a GOE, then (2)
becomes \cite{Simbel} 
\begin{equation}
P_{n}(s)=\frac{1}{n}\left[ {\text{erfc}}\left( \frac{s\sqrt{\pi }}{2n}\right) %
\right] ^{n}Q_{n}(s)\left[ \frac{\pi }{2n}s+(n-1)Q_{n}(s)\right] ,
\label{nwigner}
\end{equation}
where $Q_{n}(s)=\exp \left( -\pi ^{2}s^{2}/4n^{2}\right) /$erfc$\left( s%
\sqrt{\pi }/2n\right) $ and erfc$(x)$ is the complementary error function. \
This distribution has a shape intermediate between those of the Wigner and
the Poisson distributions. The Brody distribution \cite{brody}, 
\begin{equation}
P_{\text{B}}\left( s\right) =c_{\beta }s^{\beta }\exp \left( -\frac{c_{\beta
}}{\beta +1}s^{\beta +1}\right) \ \ \text{ with } \ \ c_{\beta }=\frac{\Gamma ^{\beta
+1}\left( 1/\left( \beta +1\right) \right) }{\beta +1},
\label{Brody}
\end{equation}
accurately reproduces the spectra of many two-dimensional systems with
mixed regular-chaotic dynamics. As we vary the parameter $\beta$ from 0 to 1 the statistics changes from  Poissonian to GOE.

The NNS distribution contains information about the spectrum in a short
range, not exceeding a maximum of three mean level spacings. \ Long range
information is provided by various higher-order correlation functions \cite
{phys.rep, mehta}. \ The most popular among these is the spectral rigidity $%
\Delta _{3}$ of Dyson and Mehta \cite{dyson}. 
Semiclassical arguments \cite{delta3} show that, for a generic integrable
system whose spectrum satisfies Poisson statistics 
\begin{equation}
\Delta _{3,\text{Poisson}}(L)=\frac{L}{15}\text{ \ \ for \ \ }L\ll L_{\text{%
max}},
\label{delta3pois}
\end{equation}
where $L_{\text{max}}=2\pi \hbar /DT_{\text{min}}$, $D$ is the mean level
spacing and $T_{\text{min}}$ is the period of the shortest classical orbit.
\ For time-reversal-invariant chaotic system the semiclassical theory gives
the following asymptotic expression 
\begin{equation}
\Delta _{3,\text{GOE}}(L)=\frac{1}{\pi ^{2}}\ln L-0.007\text{ \ \ for \ \ }%
1\ll L\ll L_{\text{max}}.
\label{delta3asy}
\end{equation}
In all cases $\Delta _{3}$ saturates to a non-universal value at $L\sim L_{%
\text{max}}$. \ The random matrix theory obtains an analytical expression
for $\Delta _{3,\text{GOE}}(L)$\ that involves double integration \cite
{phys.rep, mehta}. \ We find it more suitable to parametrise this expression
for a GOE, guided by Eq. (\ref{delta3asy}), in the form 
\begin{equation}
\Delta _{3,\text{GOE}}(L)=\frac{1}{\pi ^{2}}\left( 1-e^{-aL}\right) \left[
\ln (L+b)+c\right] ,
\label{delta3abc}
\end{equation}
with $a=8.241,b=0.944$ and $c=-0.06724$ being the best fit parameters for
values of $L$ in the range of $0\leq L\leq 50$. \ Seligman and Verbaarschot 
\cite{Seligman} suggested the following expression for the spectral rigidity
of a spectrum resulting from a random superposition of $n$ independent
sequences with fractional level densities $f_{i}$%
\begin{equation}
\Delta _{3}(L)=\sum_{i=1}^{n}\left[ \Delta _{3,i}(f_{i}L)\right] ,
\label{delta3sv}
\end{equation}
where $\Delta _{3,i}(L)$ is the spectral rigidity for the $i$th sequence. \
This proposal is not surprising because $\Delta _{3}(L)$ is essentially a
variance.

We now consider symmetry breaking in the chaotic system using the
Guhr-Weidenm\"{u}ller model\ \cite{guhr}. \ The model suggests that the
effect of the symmetry-breaking perturbation on all the unperturbed states
is equal on the average. \ This suggests defining the Hamiltonian in the
form: 
\begin{equation}
H=\left[ 
\begin{array}{cc}
H_{0} & 0 \\ 
0 & H_{0}
\end{array}
\right] +\alpha \left[ 
\begin{array}{cc}
0 & W \\ 
W^{T} & 0
\end{array}
\right] ,
\label{HW}
\end{equation}
where $H_{0}$\ represents a GOE and $W$\ is a random matrix whose elements
have Gaussian distributions with zero means and variances equal to those of
the non-diagonal elements of $H_{0}$. \ When $\alpha =0,$ the symmetry is
conserved and the Hamiltonian has the block diagonal form of Eq. (\ref{diag}) with $n=2$. \
The spectrum consists of an independent superposition of two level sequences
having equal density. \ The NNS distribution $P(s)$ is given by Eq. (\ref{nwigner}) and,
in particular, $P(0)=1/2.$ \ The spectral rigidity is given by (\ref{delta3sv}) with $%
f_{1}=f_{2}=1/2$. \ When we allow $\alpha $ to increase,\ the system
gradually evolves toward a single GOE. \ To show how this transition
proceeds, we considered an ensemble of ten 200$\times 200$ Hamiltonian
matrices of the form (\ref{HW}) for a fixed value of the perturbation strength $%
\alpha $. We numerically diagonalised these matrices. \ We obtained the NNS
distribution and spectral rigidity for each of the resulting spectra, and
measured their deviation from the GOE statistics by comparing with the Brody distribution, Eq. (\ref{Brody}),  for the NNS distribution and with the Seligman and Verbaarschot expression, Eq. (\ref{delta3sv}), for the $\Delta _{3}$\ statistics
where $\Delta _{3,1}$\ is given by the Poisson distribution (\ref{delta3pois}) and $\Delta _{3,2}$\ by the GOE (\ref{delta3abc}), so that
\begin{equation}
\Delta _{3,\text{SV}}(L)=q_{\text{SV}}\frac{L}{15}+\Delta _{3,\text{GOE}}%
\left[ (1-q_{\text{SV}})L\right] ,
\label{delta3qsv}
\end{equation}
\ We then repeated the procedure for other value of $\alpha $. \ The best
fit of the tuning parameters\ $\beta$ and $q_{SV}$ obtained for $P(s)$ and $\Delta_3(L)$ corresponding to different values of the perturbation strength $\alpha$ are shown in Fig. 1.  The error bars measure the changes of the parameters that increase the minimum $\chi^2$ values by a factor of 2. The monotonous nature of the
curves clearly shows that the transition from symmetry to no-symmetry in the
Guhr-Weidenm\"{u}ller model is gradual and one-directional.

The opposite extreme of the Guhr-Weidenm\"{u}ller model is the case when the
eigenfunctions will become either very weakly or very strongly perturbed. \
For those states in which the perturbation is weak enough, the wave function
can in principle be characterised by one of the two symmetry eigenvalues. These will be the states for which the splitting of two degenerate levels belonging to the two symmetry representations can be neglected.  The corresponding energy levels will again form two independent
sequences with fractional level densities $f_{1}=f_{2}=f/2,$ while the
levels of the strongly perturbed eigenstates will form a new sequence with
density $f_{3}=1-f$ which will grow at the expense of the other two as the
symmetry-breaking perturbation increases. \ The NNS distribution will then
be given, using (\ref{mixedseq}) and (\ref{wigner}), by 
\begin{eqnarray}
P(s) &=&\frac{\pi }{8}f^{3}se^{-\pi f^{2}s^{2}/16}{\text{erfc}}\left( \frac{%
\sqrt{\pi }}{4}fs\right) {\text{erfc}}\left[ \frac{\sqrt{\pi }}{2}\left(
1-f\right) s\right]   \nonumber \\
&&+\frac{1}{2}f^{2}e^{-\pi f^{2}s^{2}/8}{\text{erfc}}\left[ \frac{\sqrt{\pi }}{%
2}\left( 1-f\right) s\right]   \nonumber \\
&&+\frac{\pi }{2}\left( 1-f\right) ^{3}se^{-\pi (1-f)^{2}s^{2}/4}\left[ 
{\text{erfc}}\left( \frac{\sqrt{\pi }}{4}fs\right) \right] ^{2} \nonumber \\
&&+2f(1-f)e^{-\pi \left( 5f^{2}-8f+4\right) s^{2}/16}{\text{erfc}}\left( \frac{%
\sqrt{\pi }}{4}fs\right) .  
\label{threeseq}
\end{eqnarray}
for spacings greater than a certain lower limit. If we ignore the latter condition, the extrapolation of $P(s)$ into the origin yields
\begin{equation}
P(0)=f\left( 2-\frac{3}{2}f\right) ,
\label{threeseq0}
\end{equation}
We note that $P(0)$ increases with decreasing $f$ from a value of 1/2 at $f=1$, reaches a maximum
value of 2/3 at $f=2/3$, and then decreases monotonously until it vanishes at 
$f=0$. \ Equations (\ref{threeseq}) and (\ref{threeseq0}) show that the evolution of the shape of
the NNS distributions with increasing symmetry breaking, and decreasing $f$,
is not a straightforward transition from the 2-GOE behaviour to that of a
GOE. \ Indeed, by decreasing the value of $f$ we are removing some of the
levels from the initial two symmetry-invariant sequences to form the third
sequence that represents the states without symmetry. \ The distribution
will look more regular when we decrease $f$ until the level densities of the
three sequences become equal. \ If we further increase the symmetry
breaking, the initial sequences become thinner until the whole spectrum forms a single GOE sequence.

We can follow this transition better by calculating the spectral rigidity
which is, for this system, given by 
\begin{equation}
\Delta _{3}\left( L\right) =2\Delta _{3,\text{GOE}}\left( \frac{fL}{2}%
\right) +\Delta _{3\text{,GOE}}\left[ \left( 1-f\right) L\right] .
\end{equation}
In Fig. 2, the spectral rigidities corresponding to $f=1$, 0.9, 0.667, 0.333, 0.1 and 0
are given by the curves labelled by A, B, C, D, E and F, respectively, while
the curve P corresponds to a regular spectrum. \ The figure clearly shows
that $\Delta _{3}\left( L\right) $ increases as symmetry breaking increases,
and thus evolves towards the regular shape until $f$ reaches the value of
2/3. \ Only then starts $\Delta _{3}\left( L\right) $ a route leading to the
GOE curve.

\section{Comparison with acoustic resonance spectra}

In this section, we show that the proposed application of the Berry-Robnik
theory to symmetry breaking leads to a satisfactory description of the
acoustic resonance spectra of monocrystalline quartz blocks, measured by\
Ellegaard et al. \cite{ellegaard}. \ Crystalline quartz exhibits $D_{3}$
point-group symmetry about the crystal's $Z$ (optical) axis, and three
two-fold rotation symmetries about the three $X$ (piezoelectric) axes; the
latter three axes lie in a plane orthogonal to the $Z$ axis and sustain
angles of 120 degrees with respect to each other. \ Ellegaard et al. used
rectangular blocks of dimensions $14\times 25\times 40$ mm$^{3},$ cut in
such a way that all symmetries are fully broken except a two-fold ''flip''
symmetry about one of the three $X$ axes. \ A gradual breaking of this
symmetry is then achieved by removing successively bigger octants of a
sphere from one corner, thereby creating a three-dimensional Sinai billiard.
\ The acoustic spectrum is measured for 8 radii of the removed octant,
providing data of high statistical significance for the study of the gradual
breaking of a point-group symmetry.

\ Figure 3 shows the spectral rigidity for the cases when the symmetry is
present (open circles), and when it is violated by removing a tiny octant of
radius $r=0.5$ mm (closed circles). 
The straight line labelled P is for an integrable
system. We note that data for the case of tiny symmetry violation are
systematically higher than the ones for the case of conserved symmetry. \
This agrees with the prediction of the second scenario proposed above, in
which the system that undergoes a gradual symmetry breaking looks as if it
has become more regular in the early stages of the transition.

Ellegaard \textit{et al}. \cite{ellegaard} concluded from the rise of $%
\Delta _{3}$ over the theoretical prediction for a superposition of two GOE
sequences that the quartz block with conserved flip symmetry has much in
common with a pseudo-integrable system. \ They supported this conclusion\ by
means of other measurement. Pseudo-integrable (PI) systems are
non-integrable systems, yet non-chaotic. Two-dimensional PI systems,
exemplified by particles moving in a planar polygonal enclosure with
rational angles \cite{richens}, share with the integrable system that the
motion of trajectories in the phase space is restricted to two-dimensional
compact surfaces. \ But these invariant surfaces are multi-connected and not
tori as in the case of integrable systems. \ Numerical studies of the $\pi /3
$ rhombus billiard \cite{biswas, gremaud, bogomolny} show that the spectral
fluctuation properties are intermediate between those of a Poisson ensemble
and a GOE. \ Bogomolny \textit{et al}. \cite{bogomolny} studied the level
statistics of PI systems using a modified version of Dyson's stochastic
Coulomb gas model \cite{mehta}, in which the interaction between particles
is restricted only to the nearest neighbours. \ The NNS distribution obtained
in\ \cite{bogomolny} is given by 
\begin{equation}
P_{\text{PI}}(s)=4se^{-2s}.
\label{ppi}
\end{equation}
This distribution is in agreement with the numerical calculation for the
rational billiards \cite{gremaud}. \ Unfortunately, the formula obtained by
these authors for the two-point correlation function 
\begin{equation}
R_{2}(s)=1-4e^{-4s}
\end{equation}
leads to an expression for the $\Delta _{3}$ statistics that asymptotically
increases as $L/30$, which is not consistent with the results of the
numerical experiments. \ Following several authors, e.g. \cite{biswas}, we
shall represent the $\Delta _{3}$ statistics by Seligman-Verbaarschot \cite{Seligman}
formula  (\ref{delta3sv}),\ obtained for a superposition of Poisson and
GOE level sequences, 
\begin{equation}
\Delta _{3,\text{PI}}(L)=q\frac{L}{15}+\Delta _{3,\text{GOE}}\left[ (1-q)L%
\right] ,
\label{delta3pi}
\end{equation}
where $q$ is a fitting parameter. \ Biswas and Jain \cite{biswas} found $q$
to be equal to 0.2 for the $\pi /3$ rhombus billiard.\ 

We now apply the second scenario, which has been proposed in the previous
section, to the symmetry breaking in a PI system having a symmetry
represented by a quantum number that takes one of two possible values. \
When the symmetry is conserved, the spectrum is an independent superposition
of two sequences corresponding to the two representations of the symmetry. \
The NNS distribution is accordingly represented by \ (\ref{mixedseq}) and the $\Delta _{3}
$ statistics by (\ref{delta3sv}) with $n=2$, $P_{i}\left( x_{i}\right) $\ given by (\ref{ppi})
and $\Delta _{3,i}(L)$\ by (\ref{delta3pi}). \ We now again assume that as the symmetry
violation starts, the eigenfunctions of the system will be divided into two
classes. \ The first constitutes a fraction $f$ of the eigenfunctions in
which the perturbation due to symmetry breaking can be ignored. \ The levels
of these eigenstates are divided into two independent sequences exactly as
before switching the symmetry breaking on. \ The second is the class of the
strongly perturbed states that have completely lost the symmetry. \ This
latter class will here be modelled by a GOE. \ Therefore, the total spectrum
will be composed of 3 independent sequences, two with PI statistics and each
having a fractional density of $f/2$, and one with GOE statistics and
fractional density $1-f$. \ During the symmetry-breaking transition, the NNS
distribution is given by 
\begin{eqnarray}
P(s)&\! \! \! \! \! \! \! \!=\! \! \! \! \! \! \! &\frac{1}{2}f^{2}\left( 2f^{2}s^{2}+4fs+1\right) e^{-2fs}\text{erfc}%
\left[ \frac{\sqrt{\pi }}{2}(1-f)s\right] \nonumber \\
&&+\frac{1}{8}(1-f)(2+fs) \times \left[ \pi fs^{2}(1-f)^{2}+2s\left\{ \left( \pi +4\right) f^{2}-2\pi
f+\pi \right\} +8f\right] \nonumber \\
&& \times e^{-\frac{\pi }{4}(1-f)^{2}s^{2}-2fs},
\label{pfe}
\end{eqnarray}
which is obtained by substituting Eqs. (\ref{wigner}) and (\ref{ppi}) into Eq. (\ref{mixedseq}). \ The $%
\Delta _{3}$ is\ given by 
\begin{equation}
\Delta _{3}\left( L\right) =2\Delta _{3,\text{PI}}\left( \frac{fL}{2}\right)
+\Delta _{3\text{,GOE}}\left[ \left( 1-f\right) L\right] .
\label{delta3fe}
\end{equation}

We now compare our model with the acoustic resonance spectra measured by
Ellegaard \textit{et al}. \cite{ellegaard}. \ We start by estimating the
parameter $q$ in the expression (\ref{delta3pi}) for the $\Delta _{3}$ statistics of the
PI system. \ We do this by fitting the same statistics for the spectrum of the
complete quartz cube to Eq. (\ref{delta3fe}) with $f=1$ and obtain $q=0.082$. \ We then
compare Eq. (\ref{delta3fe}) with the spectral rigidities of the other spectra
considering $f$ as a free parameter. \ The best-fit values of $f$ for
spectra corresponding to different radii of the removed octants are given in
Table 1. \ We use these values to calculate the NNS distributions. \ The
results of the calculation are compared with the experimental data in Fig. 4. \
We see that the proposed model presents a satisfactory description of the
whole transition by varying a single parameter. \ The only disagreement is
between the calculated and measured values of $P(s)$ at small values of $s$ as we have already expected. The symmetry breaking decreases the probability of finding degenerate levels sharply leading to the observed dip at small $s$ in the spacing distributions \cite{ellegaard}. This dip is followed by an overshoot to restore normalisation. The width of this dip, which is about 1/10 of the mean level spacing, provides an estimate for the minimum level-splitting below which two levels can be regarded as approximately degenerate.
\

\section{Summary and conclusions}

Conventional models for symmetry breaking in chaotic systems assume that the non-diagonal matrix elements of the symmetry breaking perturbation are statistically equivalent. They predict a
monotonous transition from a mixed statistics to the GOE as one switches on
and increases the symmetry-breaking perturbation. \ This is not the behaviour
of the acoustic resonance spectra of quartz measured by Ellegaard \textit{et
al} \cite{ellegaard}. \ We propose another scenario for symmetry breaking in which the levels
are approximately separated into two groups, one with and one without
symmetry. \ This happens when the expansion of any of the wavefunctions in
terms of the eigenfunctions of the symmetry operator is either dominated by
a single term or composed of statistically equivalent contributions from all
its component. \ Considering a system in which a symmetry that has two
representations is conserved, we represent the spectrum as a superposition
of two independent sequences, one for each symmetry representation. \ Now,
as we switch on the symmetry breaking perturbation, we remove a number of
levels from these two level sequences into a third sequence. \ The latter
corresponds to the strongly perturbed states in which the symmetry has
virtually disappeared. \ Because we have increased the number of spectral
partitions, the spectrum will look as if it became more regular by the small
violation of the symmetry. \ The apparent regularity will continue until the
three sequences become equally populated. \ The further increase of symmetry
breaking will reverse the evolution of the shapes of the spectral statistics
toward the GOE shapes. \ Namely this is the behaviour found by inspection of
the acoustic resonance spectra of quartz measured by Ellegaard \textit{et al} \cite{ellegaard}%
. \ We see, for instance, from Fig. 4 that it is possible to describe the
evolution of the NNS distribution and the spectral rigidity during the
transition from a fully conserved symmetry to a fully violated one by
varying a single parameter, measuring the fractional density of states which
are practically unaffected by symmetry violation.\newpage 

\ \ \

\newpage \textbf{Table 1.} Best fit values for the fraction $f$ of states
not affected as a result of violating the point-group symmetry of a quartz
rectangular block by removing spherical octants of various radii $r$. The values of $f$ are evaluated by fitting the spectral rigidities $ \Delta _{3}\left( L\right) $ of Fig. 4. to Eq. (\ref{delta3fe}).

\begin{tabular}{|l|l|l|l|l|l|l|l|}
\hline
$r$ & 0 & 0.5 mm & 0.8 mm & 1.1 mm & 1.4 mm & 1.7 mm & huge defocusing \\ 
\hline
$f$ & 1 & 0.49 & 0.37 & 0.28 & 0.23 & 0.14 & 0 \\ \hline
\end{tabular}

\bigskip

\bigskip

{\LARGE Figure Caption}

\bigskip

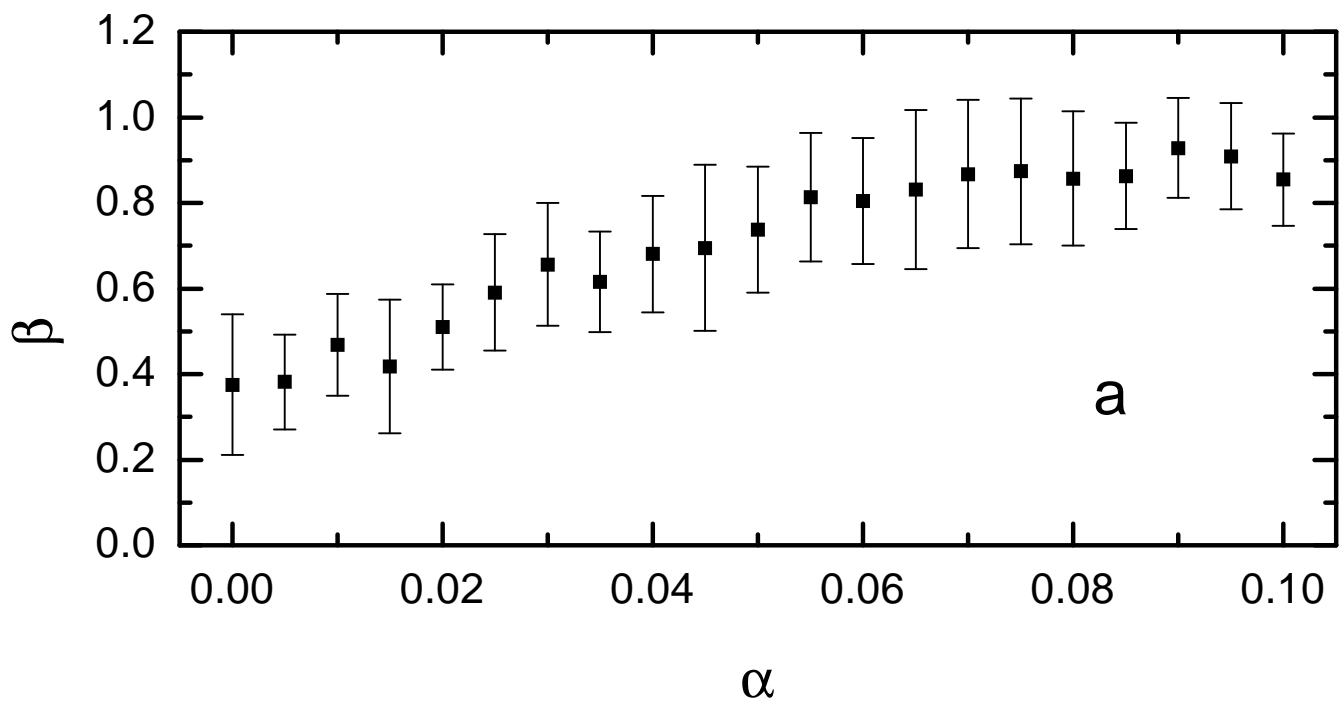
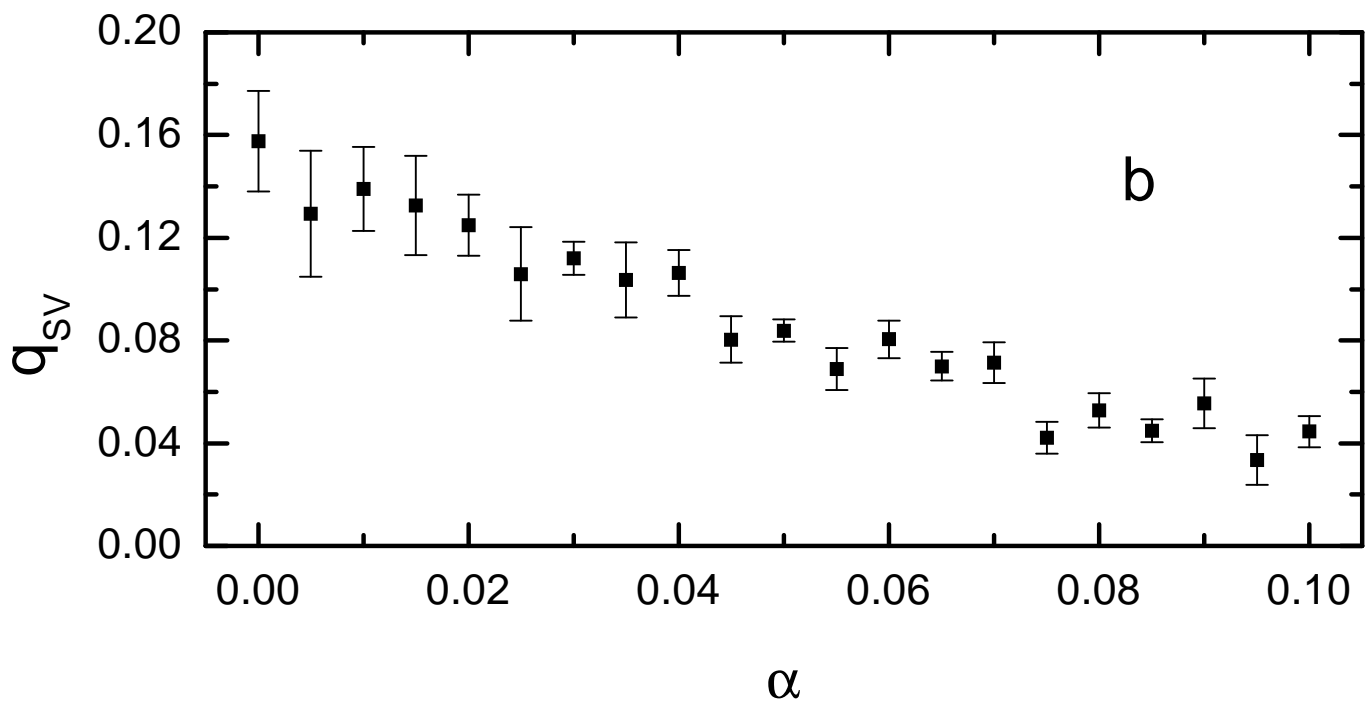
\textbf{Figure 1.} The monotonous dependence of (a) the parameter $\beta$ in the Brody distribution 
(Eq.(\ref{Brody})) for the NNS distributions $P_{\text{B}}(s)$\, and (b) the parameter $q_{SV}$ in the Seligman and Verbaarschot expression (Eq.(\ref{delta3qsv})) for the spectral rigidity $%
\Delta _{3}\left( L\right) $ that interpolate between the Poisson and GOE
statistics, obtained by a numerical experiment on a chaotic system
undergoing a gradual breaking of a symmetry according to the
Guhr-Weidenm\"{u}ller model \cite{guhr}. \ 

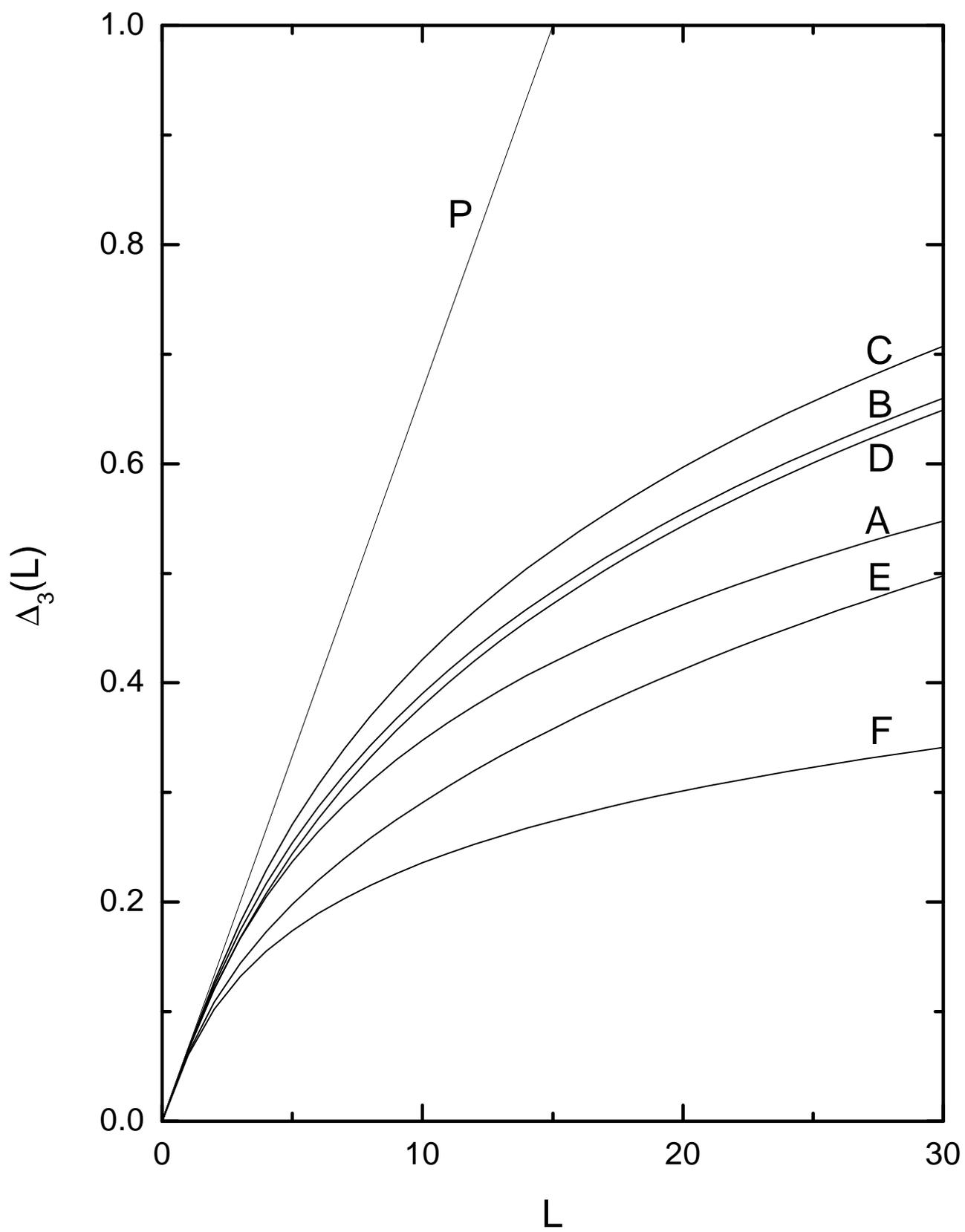
\textbf{Figure 2.} The spectral rigidity $\Delta _{3}\left( L\right) $ for a
chaotic system undergoing a gradual breaking of a symmetry according to the
scenario proposed in the text. \ The results of the calculation when the fraction
of states remaining nearly symmetry invariant $f=1$, 0.9, 0.667, 0.333, 0.1
and 0 are given by the curves labelled by A, B, C, D, E and F, respectively,
while the curve P corresponds to a regular spectrum.

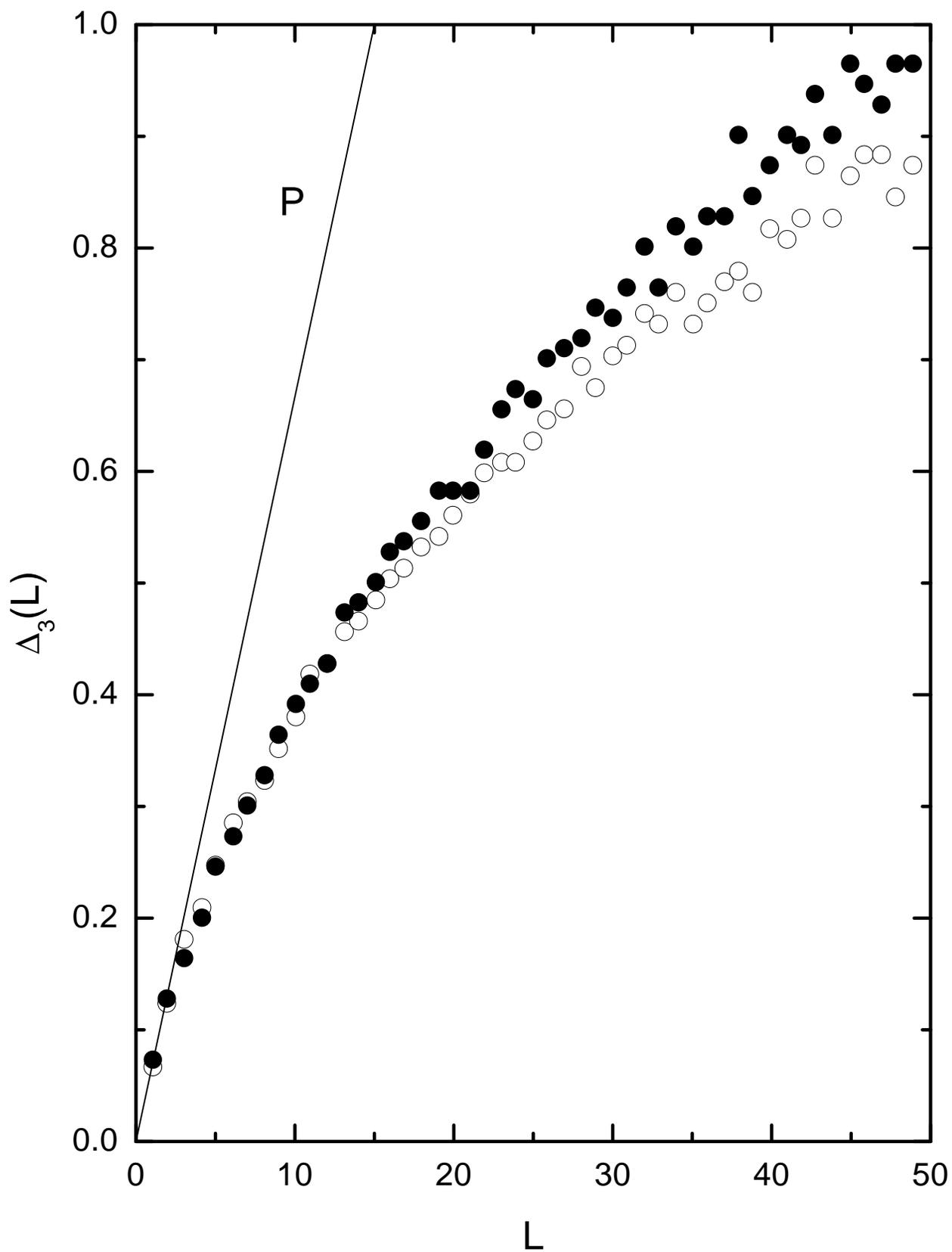
\textbf{Figure 3.} The spectral rigidity $\Delta _{3}\left( L\right) $
measured in \cite{ellegaard} for the quartz block when the symmetry is
present (open circles), and when it is violated by removing a tiny octant of
radius $r=0.5$ mm (closed circles). 
The straight line labelled P is for an integrable system.

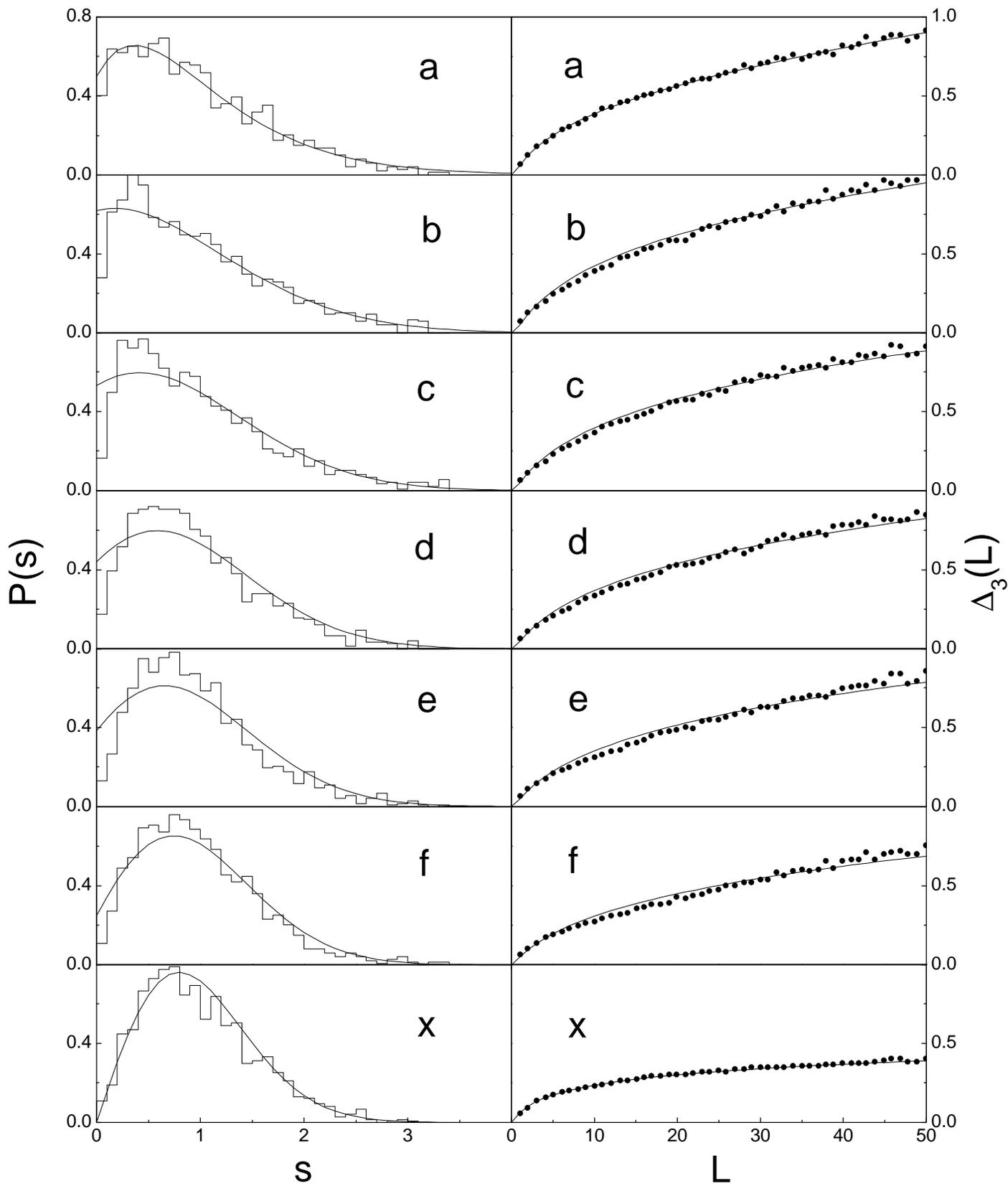
\textbf{Figure 4.} The NNS distributions $P(s)$ and spectral rigidities $%
\Delta _{3}\left( L\right) $ for the different radii of the octant removed
from a quartz block: (a) $r=0$, the flip symmetry is fully conserved, (b) $%
r=0.5$ mm, (c) $r=0.8$ mm, (d) $r=1.1$ mm, (e) $r=1.4$ mm, (f) $r=1.7$ mm,
(x) the block with the huge defocusing structure. \ The data are from Ref. 
\cite{ellegaard}, while the curves are the theoretical curves using Eqs. (\ref{pfe}) and (\ref{delta3fe}). \ Values of the parameter $f$ that measures the fractional
densities of states that keep the symmetry are given in Table 1.

\end{document}